# The Emerging Weak Antilocalization Effect in Semimetal $Ta_{0.7}Nb_{0.3}Sb_2$ Single Crystal

Lei Guo,[1] Meng Xu,[2] Lei Chen,[3] Ting Wei Chen,[3] Weiyao Zhao,[4] Xiaoling Wang,[4] Shuai Dong,[1,*] and Ren-Kui Zheng[3,5,*]

[1]School of Physics, Southeast University, Nanjing 211189, China

[2]College of Science, Hohai University, Nanjing 210098, China

[3]School of Materials Science and Engineering, Jiangxi Engineering Laboratory for Advanced Functional Thin Films, Nanchang University, Nanchang 330031, China

[4]Institute for Superconducting and Electronic Materials, & ARC Centre of Excellence in Future Low-Energy Electronics Technologies, Innovation Campus, University of Wollongong, North Wollongong NSW 2500, Australia

[5]School of Physics and Materials Science, Guangzhou University, Guangzhou 510006, China

## Abstract

Weak antilocalization (WAL) effect is commonly observed in 2D systems, or 3D topological insulators, topological semimetal systems. Here we report the clear sign of WAL effect in high quality $Ta_{0.7}Nb_{0.3}Sb_2$ single crystals, in below 50 K region. The chemical vapor transport method was employed to grow the single crystal samples, the high crystallization quality and uniform element distribution are verified by X-ray diffractions and electron microscopy techniques. Employing the Hall effect and two-band model fitting, the high carrier's mobility (> 1000 $cm^2V^{-1}s^{-1}$ in 2 – 300 K region) and off-compensation electron/hole ratio (~8:1) are obtained. Due to the different angular dependence of WAL effect and the fermiology of $Ta_{0.7}Nb_{0.3}Sb_2$ single crystal, interesting magnetic-field-induced symmetry change is observed in angular magnetoresistance. These interesting transport properties will lead to more theoretical and applicational exploration in $Ta_{0.7}Nb_{0.3}Sb_2$ and related semimetal materials.

# Ⅰ. INTRODUCTION

Benefiting from the symmetry protected nontrivial topological states exploration in condensed matter system, the transport properties of intermetallic compounds are well studied in the last decade [1–6]. Among those novel compounds, the compensated semimetals $M$Sb$_2$ ($M$=Nb, Ta) with extremely large magnetoresistance (XMR), and low temperature resistivity plateaus are attracting increasing attentions [7–14]. Theoretical and experimental results show that NbSb$_2$ and TaSb$_2$ are topological materials or good candidates, they could be weak topological insulators at zero magnetic field and can be classified as Type-II Weyl materials under the application of magnetic fields [13,14]. Due to the high chemical similarity between those two compounds, a chemical vapor transport (CVT) based high-throughput crystal growth process was reported, providing the possibility to investigate the exotic transport behavior in Ta/NbSb$_2$ system. Based on our previously transport study in Ta/NbSb$_2$ system, the carrier's compensation induced XMR effect are observed in TaSb$_2$ (or with less than 10% Nb), NbSb$_2$ (or with less than 10% Ta) crystals [12]. In Ta-Nb well mixed crystals, the magnetoresistance (MR) values at low temperatures are ~ 100% due to the off-compensated effect.

Another important transport behavior in topological related materials is weak antilocalization (WAL), which describes the antilocalization of spin resolved carrier trajectories with destructive interferences between spin-up and spin-down carriers. This ultimately reduces the resistance at low temperatures and can be destroyed by applied magnetic fields, therefore showing a cusp-like magnetic conductivity behavior [15–17]. Typically, the WAL effects are observed in the lower-dimensional systems, such as thin films or nanoflakes [18–20], due to the higher probability of scattering events occurring between two time-reversed paths, while it is relatively rare in bulk materials. A special system that exhibits WAL effect is 3D topological insulators (Bi$_2$Se$_3$, Bi$_2$Te$_3$, Bi$_2$Te$_2$Se, Bi$_2$Se$_{2.1}$Te$_{0.1}$, V,Sn:Bi$_{1.1}$Sb$_{0.9}$Te$_2$S etc.), in which the spin-moment locking of topologically protected surface states provide the source of destructive interference [21–25]. Recently WAL effect is also reported in some 3D topological semimetal crystals (LuPdBi [26,27], ScPtBi [28], YbCdGe

[29], YbCdSn [30], CaAgBi [31]), topological superconductors (LuPdBi [26,27]), and a very small number of half-Heusler compounds (ScPdBi [32], ScNiBi [33], LuNiBi [34], LuPtSb [35]). Experimentally, WAL effect can be detected measuring the magnetoconductance at low temperatures, which can be fitted via Hikami-Larkin-Nagaoka (HLN) model [36,37].

In this work, we observed clear WAL effect in CVT grown high quality single crystals $Ta_{0.7}Nb_{0.3}Sb_2$ at low temperatures. The magnetotransport property of $Ta_{0.7}Nb_{0.3}Sb_2$ crystal is similar to previously reported $TaSb_2$, $NbSb_2$ and $Ta/NbSb_2$, e.g., showing large, non-saturating MR, resistivity plateaus below ~ 10 K, and magnetic field induced resistivity upturn. The very different point is, at low temperature, the MR curves show sharp "V" shape below ~ 1 T magnetic field, and increase parabolically with magnetic fields above 1 T. The WAL shape of MR curve survives up to ~ 50 K, which can be well fitted via the HLN model. Due to the different angular dependence of WAL effect, and the fermiology of $Ta_{0.7}Nb_{0.3}Sb_2$, the symmetry of angular MR evolves with magnetic field dramatically. The WAL effect observed in $Ta_{0.7}Nb_{0.3}Sb_2$ single crystal provide a new platform to study this phenomenon, and leads to more investigation in $Ta/NbSb_2$, as well as other semimetal system.

## Ⅱ. Experimental Methods

High-quality single crystals of $Ta_{0.7}Nb_{0.3}Sb_2$ were grown using CVT method with the iodine ($I_2$) as a transport agent. Briefly, stoichiometric amount (~1.0 g) of high-purity Nb (99.95%, aladdin), Ta (99.9%, aladdin) and Sb (99.9999%, aladdin) powders, together with ~10 mg/ml iodine were sealed in a quartz tube (<$10^{-3}$ Pa) as starting materials. Next, the crystal growth was carried out in a two-zone horizontal furnace between 1000 °C (source) and 1100 °C (sink) for 7 days before cooling to room temperature. Then the rod-like shape single crystals with a typical size of 2×1×0.5 mm$^3$ and shiny surfaces [Fig. 1(a)] were successfully obtained.

The crystal orientation of the major cleavage surface was characterized by single crystal X-ray diffraction (XRD), (PANalytical X'pert equipped with Cu $K\alpha_1$ radiation). The chemical composition and the homogeneity of the single crystals were characterized via energy dispersive x-ray spectroscopy (EDS) using an x-ray energy dispersive spectrometer (Oxford Aztec X-Max80) installed on the Zeiss Supra 55 scanning electron microscope (SEM). Further, high-resolution transmission electron microscopy (HRTEM) and selected area electron diffraction (SAED) were measured using a Tecnai G2F20 S-Twin transmission electron microscope. Note that the electron beam is incident along the [1-12] crystallographic direction.

Transport measurements were performed by a physical property measurement system (PPMS-14T, Quantum Design) via standard four-probe method. Ohmic contacts were prepared on the major shiny cleavage surface using room-temperature cured silver paste. The electric current is always along the *b* axis (i.e., the [010] direction) while the direction of the magnetic field was initially along [-201] direction of the crystal during temperature and magnetic field dependent resistivity measurements and 0° of angular MR measurements.

## Ⅲ. RESULTS AND DISCUSSIONS

As shown in the insert image of Fig. 1(a), the $Ta_{0.7}Nb_{0.3}Sb_2$ crystal show rod-like shape, with length direction along the *b* axis of the crystal and the mirror-like major surface (normal vector [-201]). To check the crystal structure, XRD measurements were carried out on the shiny plane of the selected single crystals. Sharp and clean comb-like (*h0l*) (*h*=-2, -4, -6, -8; *l*=*h*/2) X-ray diffraction patterns [Fig. 1(a)] and a very small full width at half maximum (FWHM) (~0.08º) of the XRD rocking curve taken on the (-402) diffraction peak [Fig. 1(b)] fully confirmed the excellent crystalline quality of the $Ta_{0.7}Nb_{0.3}Sb_2$ single crystal. As proved before, $Ta_{0.7}Nb_{0.3}Sb_2$, as well as other $Ta/NbSb_2$ compounds share the same monoclinic structure ($C_{12/m1}$ space group) with the parent compounds $TaSb_2$ and $NbSb_2$ [7, 12, 38].

In order to accurate determine the chemical composition together with its distribution, we carried out the EDS measurement on the cleavage surface. As shown in Fig. 1(c), the Ta, Nb, and Sb element of our samples and the average chemical composition ratio was established to be 23.3%:9.9%:66.8%, which is close to the nominal stoichiometry

$Ta_{0.7}Nb_{0.3}Sb_2$. Moreover, the element mapping of the micro milled sample via EDS equipped on TEM, which aims to enhance the accuracy and reliability, also demonstrates the uniform distribution of Ta, Nb Sb elements [Figs. 1 (c)–1(g)].

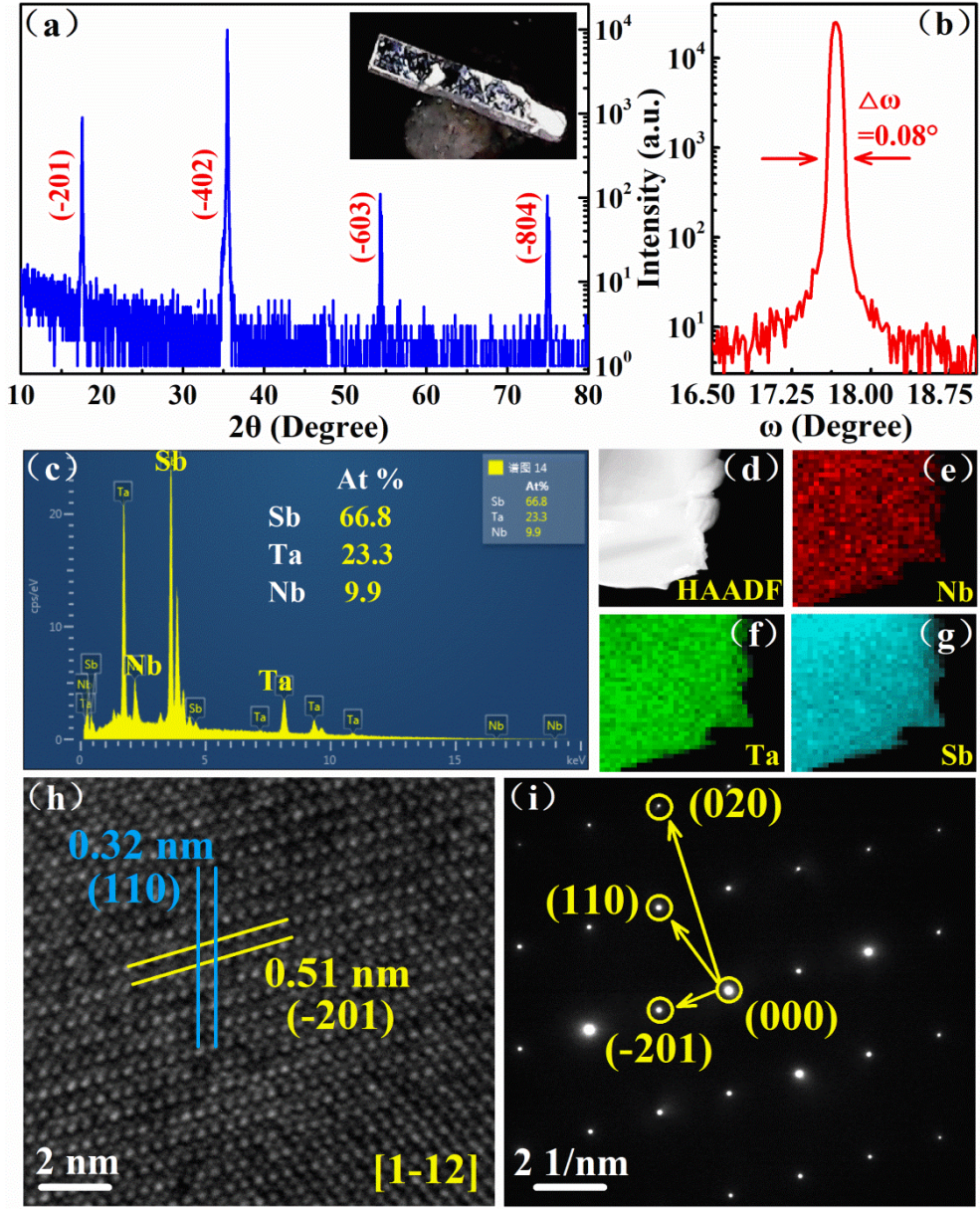

FIG. 1. (a) XRD pattern of $Ta_{0.7}Nb_{0.3}Sb_2$. Inset: an optical image of the crystal. (b) Rocking curve taken on the (-402) diffraction peak and the FWHM is 0.08º. (c) The element index in energy dispersive x-ray spectroscopy (EDS). (d-g) Element mapping patterns of $Ta_{0.7}Nb_{0.3}Sb_2$ Single crystal. (h) Atomic-resolution HRTEM image. (i) Selected area electron diffraction pattern.

Finally, the excellent crystalline quality of the single crystal is fully reflected by High-resolution transmission electron microscopy (HRTEM) result [Fig. 1(h)], which demonstrates that and its atomic sites and lattice spacing are 0.51 and 0.32 nm, corresponding to the lattice spacing of the (-201) and (110) planes, respectively. The nearly perfect microstructure naturally resulting in sharp selected area electron diffraction (SAED) pattern [Fig. 1(i)], and it can well be indexed with the $C_{12/m1}$ space group, which further confirms it possesses a monoclinic crystal structure [38]. All of these results demonstrate the high quality of the $Ta_{0.7}Nb_{0.3}Sb_2$ single crystal which provides a reliable platform to study its electronic transport properties.

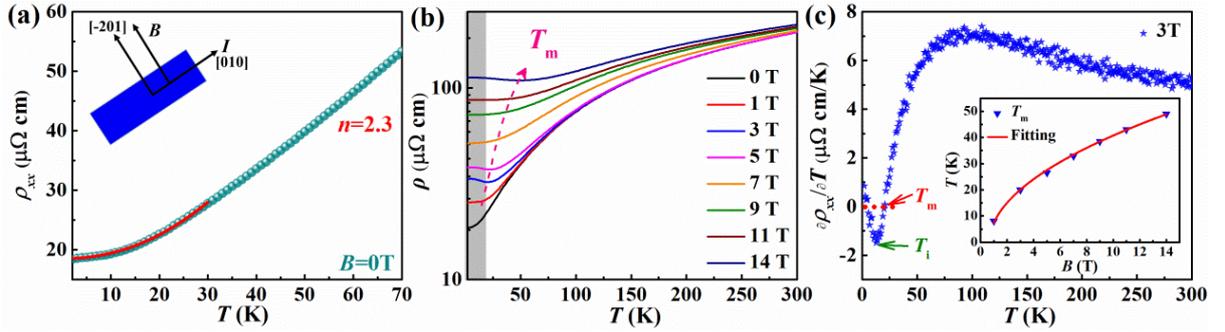

FIG. 2. (a) Zero-field resistivity (black point) and fitting results (red curve) of $Ta_{0.7}Nb_{0.3}Sb_2$ single crystal. Inset: schematic geometry of the directions of the magnetic field and the electric current. (b) Temperature dependence of resistivity in different magnetic fields, as measured using the schematic geometry shown in the inset of (a). The grey shadow areas show the resistance plateau region and the pink dash arrow indicates the resistance minimum temperature ($T_m$). (c) $d\rho_{xx}/dT$ plotted as a function of temperature, taking T = 3T as an example. Insert: $T_m$ plotted as a function of magnetic field.

As shown in Fig. 2(b), the zero-field temperature dependent resistivity (*RT*) curve of the $Ta_{0.7}Nb_{0.3}Sb_2$ crystal shows monotonically increasing within the entire temperature region (2-300 K), which suggests the metallic ground state. The resistivity at 2 K becomes as low as approximately 18 μΩ cm, which further suggesting its good metallic property and it could be mainly due to the large mean free path of charge carriers of the single crystal [39]. And the residual resistivity ratio (RRR) is approximately 10.6 under a zero magnetic field. Moreover, the resistivity in the temperature region from 2 to 30 K can be well fitted using the equation $\rho(T)=\rho_0 +AT^n$ with $n\sim2.3$, where $\rho_0$ is the residual resistivity, *A* is a constant, and *n* is a parameter indicating scattering mechanisms. Similar phenomenon has been observed in compensated semimetals of LaBi (*n* = 3) [40], LaSb (*n* = 4) [41], and Dirac semimetals of

ZrGeSe ($n$ = 3.1) [42]. This type of temperature dependence of resistivity, deviating from the pure electronic correlation-dominated scattering mechanism ($n$ = 2), can be attributed to the interband electron-phonon scattering [43].

Under the application of magnetic field ($B \geq 1$T), one can find a metal-to-insulator-like transition and resistivity plateau (the grey shadow areas) in the low-temperature region [Fig. 2(b)], both of which are sensitive and positively correlated to the magnetic field, and all these are generic features of topological semimetals [29, 30, 39–42]. To further analyze these features, we plot first derivative of resistivity ($(\partial\rho_{xx}/\partial T)$) as a function of temperature as shown in Fig. 2(c). From this figure, two characteristic temperatures can be clearly identified: 1) the metal-to-insulator transition temperature $T_m$, corresponding to $\partial\rho_{xx}/\partial T=0$ (see the red dashed guide line). 2) the onset temperature of the resistivity plateau ($T_i$), where $\partial\rho_{xx}/\partial T$ shows the minimum (see the olive solid arrow), remains slight changed with increasing magnetic fields. The obtained $T_m$ value increases monotonically with increasing magnetic fields, as shown in the insert of Fig. 2(c), and the relationship between $T_m$ and $B$ can be described by the equation $T_m \sim a\,(B-B_0)^{1/\nu}$. The parameter $\nu$ is ~1.96, which is close to that of compensated semimetals such as Bi ($\nu$=2) [44], WTe$_2$ ($\nu$=2) [45]. In Ta/NbSb$_2$ research, one should note that, the carriers are off-compensation within 10% < Nb percentage < 50% region, and thus the magnetic field induced resistivity upturn effects are not as strong as in TaSb$_2$ [12]. Here, in Ta$_{0.7}$Nb$_{0.3}$Sb$_2$ crystal, the low temperature upturn effect is relatively (compared with TaSb$_2$ or NbSb$_2$) weak [7, 9, 10], which may be due to the off-compensation electron and hole carriers [12].

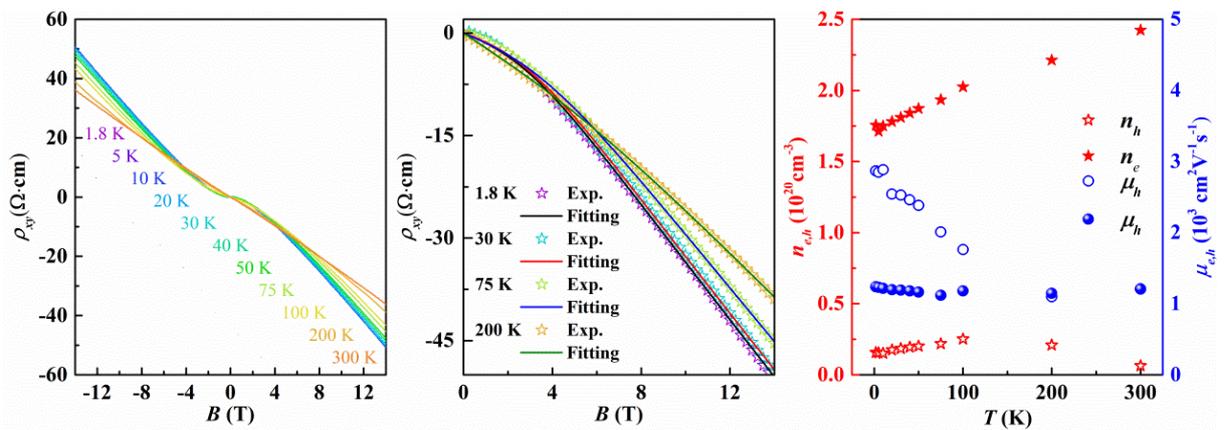

FIG. 3. (a) Hall resistance as a function of the magnetic field at different fixed temperatures for the Ta$_{0.7}$Nb$_{0.3}$Sb$_2$ single crystal. (b) Double-band model fitting for Hall curves, in which the representative data at 1.8, 30, 75 and 200 K are well fitted. (b) The carrier's density and mobility obtained via double-band fitting, plot at a function of temperatures.

To further understand the carrier's properties in the Ta$_{0.7}$Nb$_{0.3}$Sb$_2$ crystal, we performed Hall resistivity ($\rho_{xy}$) measurements at different fixed temperatures ranging from 1.8 to 300 K, which are plotted in Fig. 3 (a). The nonlinear tendency of the $\rho_{xy}$ versus $B$ curves suggests the existing of multi-type carriers, especially at low temperature and low field region. Here, we consider the semiclassical two band model to analyze its carrier's density & Hall mobility [43, 46]. The total conductivity tensor σ can be expressed in the complex representation,

$$\hat{\sigma} = e\left[\frac{\mu_e n_e}{1+i u_e B} + \frac{\mu_h n_h}{1-i u_h B}\right] \qquad (1)$$

where $n$ and $\mu$ are the carrier density and carrier mobility, respectively. The subscript $e$ and $h$ denote electron and hole, respectively. Since $\rho = \sigma^{-1}$ the conductivity tensor σ in Eq. (1) can be transformed to resistivity tensor $\rho$ whose real and imaginary parts $\rho_{xx}$ and $\rho_{xy}$, respectively, can be written as:

$$\rho_{xx}(B) = \text{Re}(\rho) = \frac{1}{e}\frac{(n_h\mu_h+n_e\mu_e)+(n_h\mu_e+n_e\mu_h)\mu_e\mu_h B^2}{(n_h\mu_h+n_e\mu_e)^2+(n_h-n_e)^2\mu_e^2\mu_h^2 B^2} \qquad (2)$$

$$\rho_{xy}(B) = \text{Im}(\rho) = \frac{B}{e}\frac{(n_h\mu_h^2-n_e\mu_e^2)+(n_h-n_e)\mu_e^2\mu_h^2 B^2}{(n_h\mu_h+n_e\mu_e)^2+(n_h-n_e)^2\mu_e^2\mu_h^2 B^2} \qquad (3)$$

We fitted the $\rho_{xy}$ data using Eq. (3) [Inset of Fig. 3(a)] and obtained the fitting parameters $n_e$, $n_h$, $\mu_e$, $\mu_h$, which are the carrier density and mobility of electrons and holes, respectively. In order to demonstrate the fitting clearly, the inset figure only shows 4 different temperature curves at the positive magnetic field region as representative. The fitting parameters as a function of temperature are displayed in Fig. 3(b). It is evident that the electron and hole mobility $\mu_e$ and $\mu_h$ decrease with increasing temperature while the electron and hole carrier densities are the opposite. At 300 K, the Hall curve is nearly a straight line with negative slope, where the hole carriers' density is ~ 0. The negative temperature variation of mobility reflecting that the lattice scattering scatting dominates the temperature-dependent mobility. Overall, there are 2 important features could be found from Hall measurements: 1) the ratio $n_e/n_h$ is about 8, which verifies the off-compensation electron and hole carriers. 2) even at room temperature, the carrier mobility still keeps a relatively high value of ~$1.1\times 10^3$ cm$^2$V$^{-1}$s$^{-1}$, which could contribute to the large magnetoresistance (MR) effect. The observation of high mobility over a wide temperature region up to room temperature in Ta$_{0.7}$Nb$_{0.3}$Sb$_2$ single crystal may make it a promising system of high efficiency electronic devices.

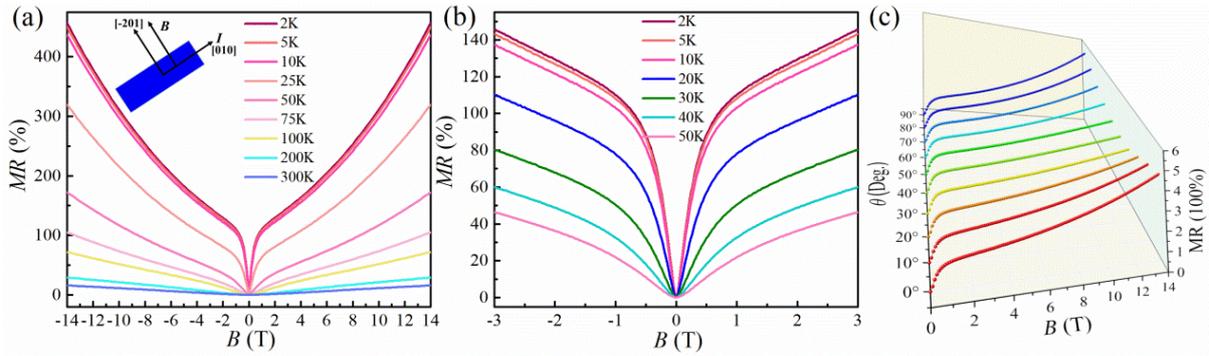

FIG. 4. (a) Magnetoresistance plotted as a function of magnetic field at different fixed temperatures for the Ta$_{0.7}$Nb$_{0.3}$Sb$_2$ single crystal. (b) A zoom-in MR measurement below 50 K ranging from -3 T to 3 T. (c) Resistivity plotted as a function of the magnetic field at various angles at $T=2$ K.

Due to the high carrier's mobility, a sizeable MR effect is expected in Ta$_{0.7}$Nb$_{0.3}$Sb$_2$ crystal. Transverse MR describes the change of resistance with the magnetic field applied perpendicular to the electric current, which is defined by $\mathrm{MR} = \frac{\rho(B)-i(0)}{\rho(0)}$, where $\rho(B)$ is the resistivity in an applied magnetic field $B$. The MR measurements at several fixed temperatures ranging from 2 to 300 K are shown in Fig. 4(a). First, it can be seen that the MR of Ta$_{0.7}$Nb$_{0.3}$Sb$_2$ crystal shows quasi-parabolic behaviors with no sign of saturation up to 14 T with increasing magnetic field and the MR versus $B$ curves almost coincide with each below

10 K, which agrees with the existence of aforementioned resistivity plateaus. For $T$=2 K and $B$=14 T, MR is ~ 457 %, which is 1-2 orders of magnitude lower than that of the parent compounds, but also can comparable to the values of many previously reported semimetals [27–30]. After Ta/Nb doping, its inter-valley scattering may enhance due to the increased atomic disorder, together with its Fermi level shifts away from the compensation area, which finally resulting in the decrease in MR [12]. For $T \geq 100$ K, MR decreases remarkably with the increasing temperature and finally reaches 15.7 % at $T$=300 K, $B$=14 T, which is quite similar to that observed in traditional metal materials.

Most interestingly, in the low-field region (from -0.5 T to 0.5 T, below 50 K), MR shows a sharp cusp-like feature which resembles the WAL effect [15–17,47]. In order to better observe this feature, we made more precise measurements in a relatively small range, as shown in Fig. 4(b). It is obvious that MR increases sharply in the low field region below 50 K, and a large MR of 120% is obtained in a magnetic field of 1 T at 2 K. This large and highly sensitive low-field MR suggests that the $Ta_{0.7}Nb_{0.3}Sb_2$ could be potential candidate for highly sensitive sensors. Fig. 4c demonstrates the MR curves at 2 K, with magnetic field tilting from [-201] direction (always perpendicular to the current during rotation), in which the WAL effect can be observed at all tilting angles. Another point is, during rotating, the MR 14 T MR value first decreases to a minimum value at ~ 50°, then increases. To further study this anisotropy effect, angular dependent MR is employed, and will be discussed later.

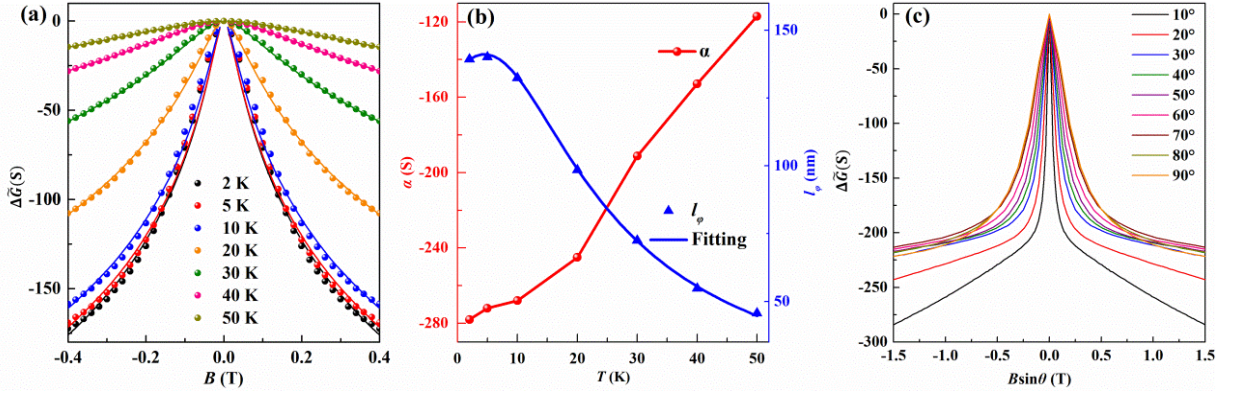

FIG. 5. (a) Magnetoconductance plot ΔG as a function of magnetic field. The solid lines are the fitting results via HLN model. (b) Temperature dependence of the α parameter, and the phase coherence length $L\varphi$ as a function of temperature, extracted from the HLN model. The solid lines are fits to equation (2).

As mentioned above that the MR shows a prominent cusp in the low-field region below 50 K which is the reminiscent of the weak antilocalization (WAL) effect and is generally found in three-dimensional (3D) systems with strong spin-orbital coupling (SOC) or two-dimensional (2D) nontrivial conducting surface states of topological insulators [18–35]. Recently WAL effect has also been reported in some 3D semimetal crystals, such as ScPtBi, YbCdGe, YbCdSn, CaAgBi, ScPdBi) [28–32]. To investigate the origin of the WAL effect, the Hikami-Larkin-Nagaoka (HLN) model is applied to analyze the 2D magnetoconductivity which can be expressed as follows:

$$\Delta G(B) = G(B) - G(0) = \frac{\alpha e^2}{2\pi^2 \hbar}\left[\Psi\left(\frac{1}{2} + \frac{\hbar}{4 e l_\phi^2 B}\right) - \ln\left(\frac{\hbar}{4 e l_\phi^2 B}\right)\right] \quad (1)$$

where G(B) and G(0) are the conductance at finite magnetic field and at zero magnetic field, Ψ is the digamma function, $l_\phi$ is the phase coherence length, and α =−1/2 per conduction channel, respectively [36–37]. Thus, the number of conducting channels can be estimated from the value of α. The WAL is general observed in 2D-topological insulators and the HLN model can be applied directly to these 2D system. The parent compounds $Ta/NbSb_2$ has been predicted to be weak topological insulators [7–14], so it makes sense to study the origin of the

WAL effect by the HLN model. However, this analysis cannot be applied directly to the conductivity data of the metallic-like 3D system. In these cases, the change of the magnetoconductivity with the applied perpendicular magnetic field is by a factor of up to at least 100 times too large, so that the value of α can not be obtained in a valid range within the HLN model [22]. Therefore, the magnetoresistance to be fitted by the HLN formula is required to be divided by the number of contributing 2D layers $Z^*$, which is about half the number of quantum layers, i.e. $Z^* = 0.5 \times t/(1 \text{ nm})$. Hence, we define the magnetoconductance per contributing 2D layer as: $\Delta \widetilde{G}(B) = \Delta G(B)/ Z^*$, and fit $\Delta \widetilde{G}(B)$ in terms of HLN formula [22, 24]. As shown in Fig. 5(a), the symbols represent the experimental $\Delta \widetilde{G}(B)$ data in the low-field region below 50 K and the corresponding lines are the fitting curves using HLN formula. The agreement between the theoretical fitting and the experimental is excellent, indicating the magnetoresistance behaviors in accordance with the HLN model. The variation of fitting parameters α and $l_\phi$ with temperature is shown in Fig. 5(b). The value of α is in the order of $10^2$, which is much larger than that of 2D systems. Although such large *α* value is rare, similar results are also observed in several 3D topological materials recently, such as LuPdBi, ScPtBi, YbCdGe, YbCdSn, CaAgBi and ScPdBi [27–32], which can be attributed to the dominance of conducting channels in the bulk, implying that the WAL effect in the $Ta_{0.7}Nb_{0.3}Sb_2$ crystal mainly originates from the the 3D bulk contribution. The value of α remains almost constant for temperatures up to 10 K, which indicates that the number of conducting channels is independent of temperature at very low temperature. As the temperature increases further, the value of α decreases dramatically, hinting the WAL effect weakens with increasing temperature, which corresponds to the resistance change in Fig. 5(a).

On the other hand, considering the electron–electron interaction and the electron–phonon scattering, the phase coherence length $l_\phi$ as a function of temperature can be fitted by the following formula:

$$\frac{1}{l_\phi^2(T)} = \frac{1}{l_\phi^2(0)} + A_{ee}T + A_{ep}T^2 \tag{2}$$

Here, $l_\phi(0)$ is the zero-temperature phase coherence length, and $A_{ee}T$ and $A_{ep}T^2$ represent the electron-electron and electron-phonon interaction, respectively. In Fig. 1(d), the symbols represent the original value of $l_\phi(T)$ obtained from the fitting of Eq. (1) and the solid lines show the fittings curves of $l_\phi(T)$ vs. $T$ using equation (2). The best fit yields $l_\phi(0) = 136\ nm, A_{ee} = -1.79 \times 10^{-6}\ (nm^2K)^{-1},\ A_{ep} = 2.13 \times 10^{-7}\ (nm^2K^2)^{-1}$, which indicates the predominant scattering mechanism is the electron-electron scattering.

To further confirm the origin of the WAL effect in the $Ta_{0.7}Nb_{0.3}Sb_2$ single crystal, we carried out the MR measurements and calculated the corresponding $\Delta\widetilde{G}$ by varying the magnetic field direction with respect to the current at 2K. The variation of $\Delta\widetilde{G}$ is displayed in Fig. 5(c) as a function of the normal component of the applied magnetic field $B\sin\theta$, where $\theta$ is the angle between the magnetic field and the current. If the WAL is caused by 2D surface states, the $\Delta\widetilde{G}$ vs $B\sin\theta$ curves should merge on to a universal curve, suggesting that the magnetoconductance depends mainly on the perpendicular component of the applied magnetic field [21,24,29]. If the WAL originates from the strong SOC in 3D bulk, the magnetoconductance should be independent of the tilt angles of the magnetic field [32–35]. As shown in Fig. 5(b), the nine curves with different tilt angles $\theta$ deviate from each other, confirming that the WAL effect in the bulk single crystal of $Ta_{0.7}Nb_{0.3}Sb_2$ mainly results from the contribution of the 3D bulk strong spin-orbit coupling rather than the 2D surface states.

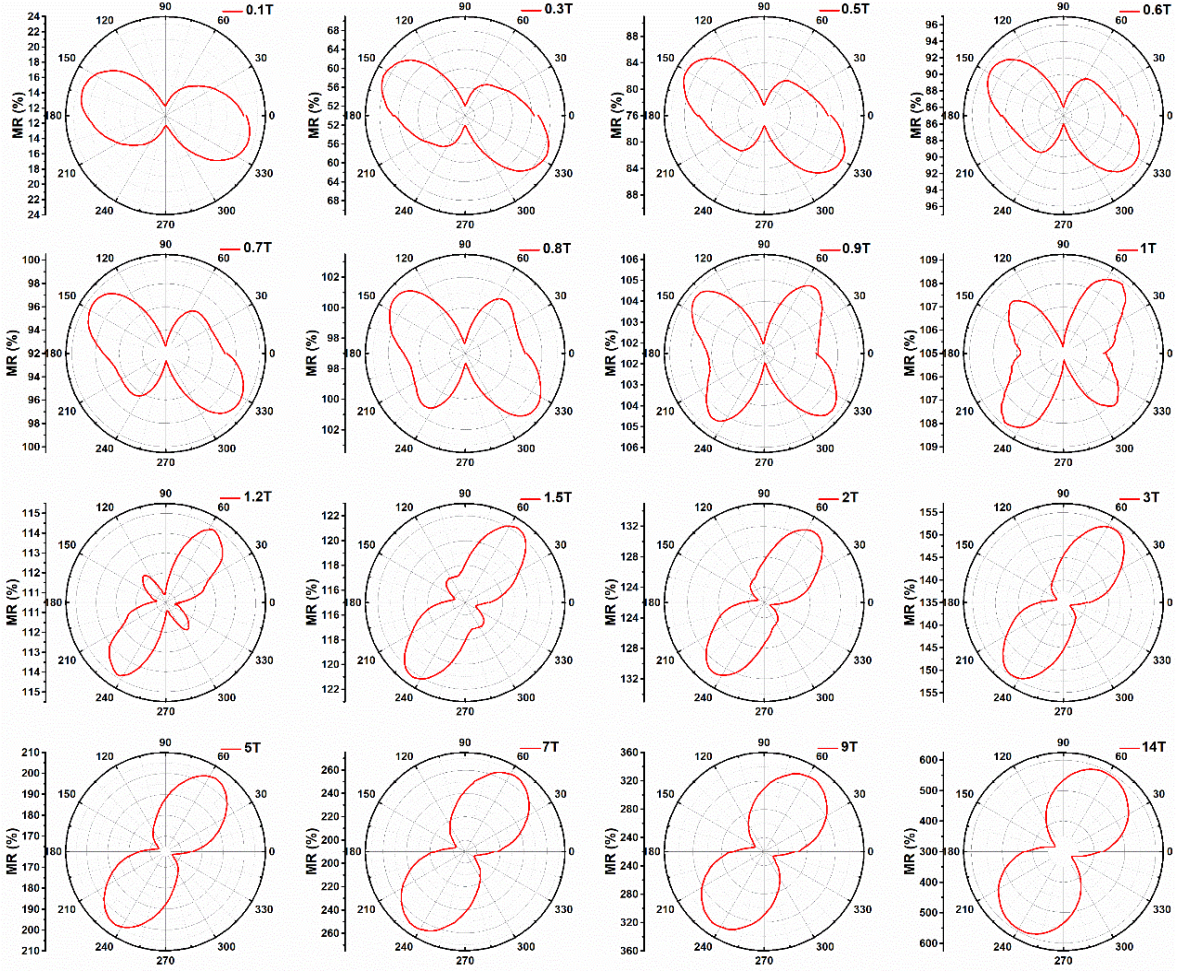

FIG. 6. The Angular dependent MR curve at 2 K and various magnetic fields.

Figures 4(c) and 5(c) demonstrate the 2K MR curves at different position, in which the anisotropy of MR at different fields can be observed. To show this anisotropic MR behavior clearer, we conducted angular dependent MR (AMR) measurement at 2 K and various magnetic fields, as shown in Fig. 6. Note that, the starting point (0°) is magnetic field along [-201], and keeping perpendicular to the current (along [010] direction). The rotation is asymmetric near 0°, where the positive direction rotates towards [001] direction, and negative rotation direction is to [100] direction. In low field region, where the WAL dominant the MR behavior, the maximum MR is obtained at ~ -35°, which is roughly the [100] axis. An interesting point is, the minimum MR is observed at 90°, not the [001] axis, therefore resulting a twisted figure of "8" shape AMR. At high magnetic fields (> 3 T), the maximum of

MR can be found along [001] direction, and the minimum is along [100] direction, thus resulting a figure of "8" shape AMR. Both in low field limits and high field limits, the AMR are two-fold symmetry or roughly two-fold symmetry, however, with different maximum MR angle. Therefore, during increasing the magnetic field, the maximum MR angle rotating is expected. As shown in Fig. 6, between 0.8 and 1.2 T, the AMR show roughly four-fold symmetry, which probably originates from the competition of WAL effect in low fields, and fermiology dominance in high fields.

## IV. CONCLUSIONS

In this letter, the magnetotransport properties of high quality CVT grown $Ta_{0.7}Nb_{0.3}Sb_2$ single crystal are investigated. The single crystal XRD, SEM, TEM techniques are employed to verify the crystallographic quality and uniformly element distribution. $Ta_{0.7}Nb_{0.3}Sb_2$ shows semi-metallic transport behavior, with large MR at low temperatures. The resistivity plateaus and magnetic field induced resistivity upturn with cooling are also observed in such a semimetal single crystal. To figure out the carrier's properties, Hall measurements and two-band model fitting are employed. The off-compensation electron and hole carrier densities are roughly 8 (electrons) :1 (holes) at low temperatures, with high mobility (above 1000 $cm^2V^{-1}s^{-1}$) in 2 – 300 K region. The interesting point of $Ta_{0.7}Nb_{0.3}Sb_2$ single crystal is that, the WAL effect can be observed at below 50 K region, which is probably induced by the electron-electron correlation effect. Due to the different angular dependence of WAL effect, and the fermiology in $Ta_{0.7}Nb_{0.3}Sb_2$ single crystal, the AMR symmetry changes roughly from [100]-maximum two-fold symmetry (< 0.6 T) to four-fold symmetry (0.8 – 1.2 T), to [001]-maximum two-fold symmetry (> 3 T). The combination of high carrier's mobility and interesting WAL effect in $Ta_{0.7}Nb_{0.3}Sb_2$ single crystal will attract more theoretical and applicational study in $Ta/NbSb_2$ and other semimetal materials.

## ACKNOWLEDGMENTS

WZ and XW acknowledge the support from ARC Centre of Excellence in Future Low-Energy Electronic Technologies (CE170100039).